\documentclass[aps,prl,reprint,balancelastpage,showpacs,floatfix]{revtex4-1}
\usepackage{amsmath,,amssymb}
\usepackage{graphicx}
\usepackage{braket}
\begin{document}
\title{Antiferromagnetic spin correlation of SU($\mathcal{N}$) Fermi gas in an optical super-lattice}
\author{Hideki Ozawa}
\altaffiliation{Electronic address: hideki$\_$ozawa@scphys.kyoto-u.ac.jp}
\affiliation{Department of Physics, Graduate School of Science, Kyoto University, Kyoto 606-8502, Japan}
\author{Shintaro Taie}
\affiliation{Department of Physics, Graduate School of Science, Kyoto University, Kyoto 606-8502, Japan}
\author{Yosuke Takasu}
\affiliation{Department of Physics, Graduate School of Science, Kyoto University, Kyoto 606-8502, Japan}
\author{Yoshiro Takahashi}
\affiliation{Department of Physics, Graduate School of Science, Kyoto University, Kyoto 606-8502, Japan}
\date{\today}

\begin{abstract}
Large-spin cold atomic systems can exhibit unique phenomena that do not appear in spin-1/2 systems. We report the observation of nearest-neighbor antiferromagnetic spin correlations of a Fermi gas with SU($\mathcal{N}$) symmetry trapped in an optical lattice. The precise control of the spin degrees of freedom provided by an optical pumping technique enables us a straightforward comparison between the cases of SU(2) and SU(4). Our important finding is that the antiferromagnetic correlation is enhanced for the SU(4)-spin system compared with SU(2) as a consequence of a Pomeranchuk cooling effect. This work is an important step towards the realization of novel SU($\mathcal{N}>2$) quantum magnetism.
\end{abstract}
\maketitle
%
%
Strongly correlated fermionic many-body systems play a fundamental role in modern condensed-matter physics. A central model for these systems is the Fermi-Hubbard model (FHM), originally developed for describing interacting electrons in a crystal. For a strong repulsive interaction, the two-component or SU(2) FHM is known to give rise to a paramagnetic Mott insulator at a higher temperature, whereas an antiferromagnetic order emerges below the N\'{e}el temperature \cite{auerbach2012interacting}. In spite of intensive study for the FHM, reaching a complete understanding remained an elusive task, even for the 1/2 spin case. The development of experimental implementation of the FHM with ultracold fermionic atoms in optical lattices has provided a new approach for advancing our understanding of strongly correlated fermions \cite{Bloch2012}. The high controllability and simplicity of these systems allow systematic study over an extremely wide range of system parameters. The milestone experiments in the strongly correlated regime are recently reported realization of an antiferromagnetic correlation and order for two-component atoms in optical lattices \cite{Greif1307, PhysRevLett.115.260401, Hart2015, Boll1257, Parsons1253, Cheuk2016, Mazurenko2017, Brown1385}.

While a great deal of progress has been made for two-component fermionic atoms, many body physics for multi-component fermionic atoms is hardly explored despite the theoretical interest \cite{PhysRevLett.92.170403, PhysRevLett.99.130406, PhysRevLett.103.135301, Gorshkov2010, Capponi2016}. Many theories have predicted that the multi-component fermionic system should exhibit rich and exotic orders at low temperatures. Fermionic isotopes of alkaline-earth-like atoms, such as ytterbium ($^{173}$Yb  \cite{PhysRevLett.98.030401}) and strontium ($^{87}$Sr \cite{PhysRevLett.105.030402, PhysRevA.82.011608})  in a quantum degenerate regime are suitable for this aim owing to their SU($\mathcal{N}=2I+1$) symmetric repulsive interactions for nuclear spin $I$ \cite{PhysRevLett.91.186402, 1367-2630-11-10-103033, Gorshkov2010}, allowing us to access the SU($\mathcal{N}>2$) FHM. The realization of SU(6) Mott insulating phase with $^{173}$Yb ($I=5/2$) atoms in an optical cubic lattice opens up the door of this direction of the research \cite{Taie2012, PhysRevX.6.021030}. Yet, quantum magnetism with SU($\mathcal{N}$) symmetry has not been achieved due to the required low temperature.

In this work, we measure and analyze the antiferromagnetic spin correlation of SU($\mathcal{N}=4,2$) Fermi gas of $^{173}$Yb in an optical dimerized cubic lattice (Fig.\ref{image1}). This system is described by the SU($\mathcal{N}$) FHM in a dimerized lattice as
\begin{eqnarray}
\hat{H}_{\rm{FH}}&=& \hat{H}_0 + \hat{H}_t, \label{eq:FHM}\\
\hat{H}_0&=&-t_{\rm{d}} \sum_{\left<i,j \right>_{\mbox{\bf{-}}}^{\rm{x}}, \sigma} \left( \hat{c}_{i,\sigma}^{\dagger} \hat{c}_{j,\sigma} + {\rm{H.c.}} \right) \nonumber \\
&& + \frac{U}{2} \sum_{i, \sigma \neq \sigma^{\prime}} \hat{n}_{i, \sigma} \hat{n}_{j, \sigma^{\prime}} - \mu \sum_{i,\sigma} \hat{n}_{i, \sigma}, \\
\hat{H}_t&=&-t \sum_{\left<i,j \right>_{-}^{\rm{x}}, \sigma} \left( \hat{c}_{i,\sigma}^{\dagger} \hat{c}_{j, \sigma} + {\rm{H.c.}}  \right) \nonumber \\
&& -t_{\rm{yz}} \sum_{\left< i,j \right>_{-}^{\rm{yz}}, \sigma} \left( \hat{c}_{i,\sigma}^{\dagger} \hat{c}_{j, \sigma} + {\rm{H.c.}}  \right),
\end{eqnarray}
where $\hat{c}_{i,\sigma}$ is the fermionic annihilation operator for a site $i$ and spin $\sigma$, $\hat{n}_{i,\sigma}=\hat{c}_{i,\sigma}^{\dagger} \hat{c}_{i,\sigma}$ is the number operator, $U$ is the on-site interaction energy, $\mu$ is the chemical potential, and $t_d, t, t_{\rm{yz}}$ are the tunneling amplitudes between the nearest neighbors in the strong link $\left<i,j \right>_{\mbox{\bf{-}}}^{\rm{x}}$, the weak link $\left<i,j \right>_{-}^{\rm{x}}$ along the $x$ axis, and the weak link $\left<i,j \right>_{-}^{\rm{yz}}$ along the other two axes, respectively. 
To reach the regime of quantum magnetism, we strongly dimerize the cubic lattice along the $x$ direction, where the exchange interaction energy within the dimer is enhanced. As a result, we observe an excess of singlets compared with triplets. By developing a technique for optically inducing a singlet-triplet oscillation (STO) \cite{PhysRevLett.105.265303} with an effectively produced spin-dependent gradient, the realization of the antiferromagnetic correlation is confirmed. We investigate the spin correlation of the SU(4) system in comparison with SU(2) over a wide range of entropy. This work demonstrates the important role of large spin degrees of freedom on the quantum magnetism.

%
%
\begin{figure}[tb]
\begin{center}
\includegraphics[width=85mm]{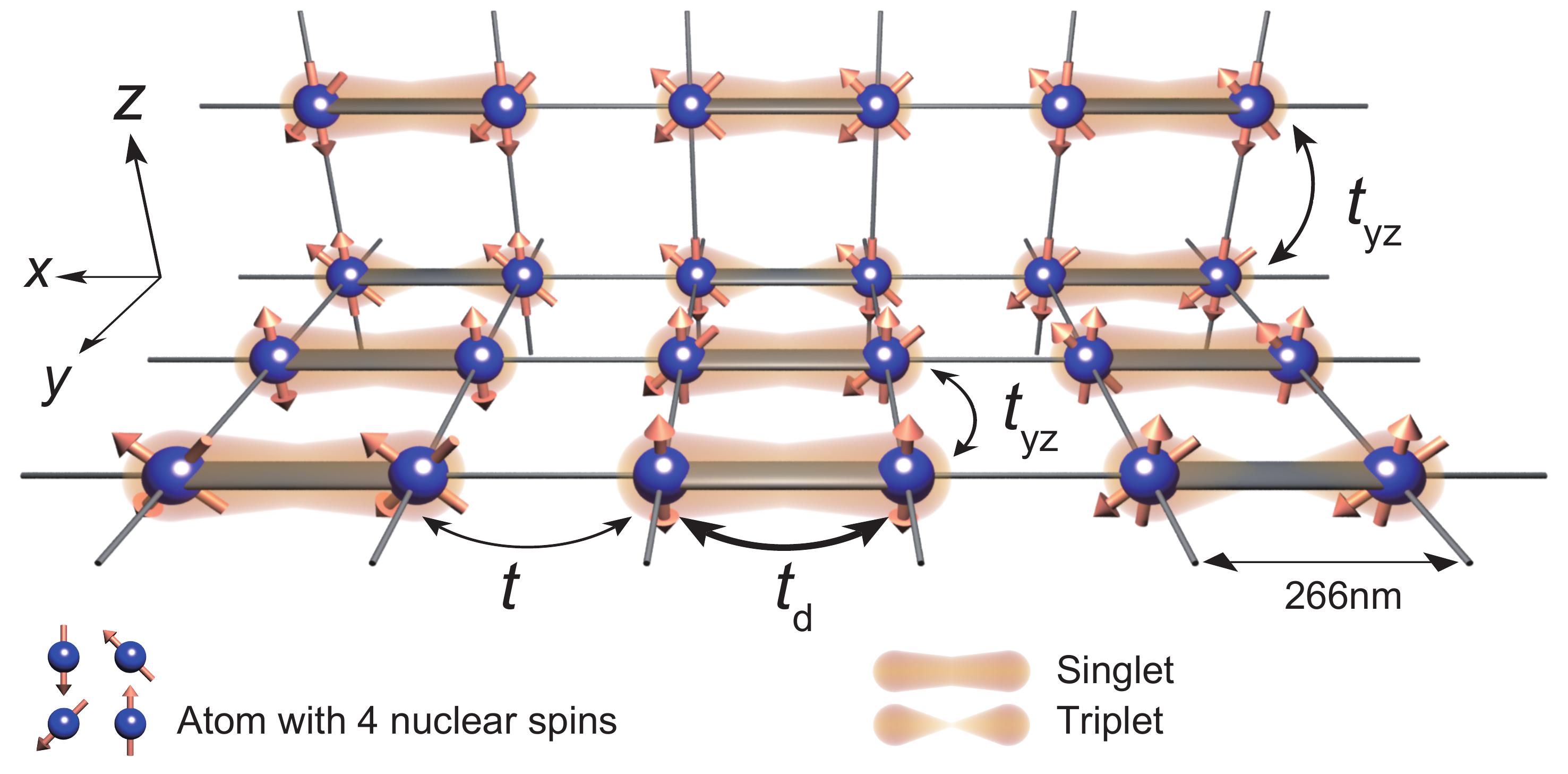}
\caption{Schematic view of the nearest-neighbor spin correlations in a four-component mixture of fermionic atoms prepared in a dimerized cubic lattice with the strong intra-dimer tunneling $t_{\rm{d}}$ and weak inter-dimer tunnelings $t, t_{\rm{yz}}$.}
\label{image1}
\end{center}
\end{figure}
We begin with describing our experimental setup. A sample is prepared by loading an evaporatively cooled two- or four-component Fermi gas of $^{173}$Yb into an optical superlattice with a dimerized cubic geometry. Our optical dimerized lattice potential is given by
\begin{eqnarray}
V(x,y,z) &=& - V_{\rm{short}}^{(x)} {\rm{cos}}^2 (2k_{\rm{L}}x+\pi /2)- V_{\rm{long}}^{(x)} {\rm{cos}}^2 (k_{\rm{L}}x) \nonumber \\
& &  -V_{\rm{short}}^{(y)} {\rm{cos}}^2(2k_{\rm{L}} y)- V_{\rm{short}}^{(z)} {\rm{cos}}^2 (2k_{\rm{L}} z), \label{eq:potential}
\end{eqnarray}
where $k_{\rm{L}}=2\pi/\lambda$ is a wave number of a long lattice, for which we choose $\lambda=1064$ nm. The short term stability of the relative phase between short and long lattices along $x$-axis is $\pm0.001\pi$ according to the relative laser linewidth. The typical phase drift is $\pm0.01\pi$ per day. All measurements of sequential data set were finished within 1 hour of the last phase calibration. In the following, we specify each lattice depth as $s_{\rm{L}}=[ (s_{\rm{long}}^x, s_{\rm{short}}^x), s_{\rm{short}}^y, s_{\rm{short}}^z ]=[ (V_{\rm{long}}^{(x)}, V_{\rm{short}}^{(x)}), V_{\rm{short}}^{(y)}, V_{\rm{short}}^{(z)} ]/E_{\rm{R}}$, where $E_{\rm{R}}= \hbar^2 k_{\rm{L}}^2/(2m)^2$ is the recoil energy for the long lattice. Unless mentioned, atoms are initially loaded into the lattice depth of $s_{\rm{L}}=[(20,20.8),48,48]$, which corresponds to the Hubbard parameters of $U/h=3.0$ kHz, $t_{\rm{d}}/h=1.0$ kHz, $t/h=37$ Hz, and $t_{\rm{yz}}/t=1.3$. The tunnelings along the dimerized lattice are determined by fitting a tight-binding model to the bands of the first principle calculation. For the on-site interaction, we constructed the Wannier function with the method described in Ref.\cite{PhysRevA.90.013614}. We also estimated the beyond-Hubbard terms such as nearest-neighbor interaction and density-induced tunneling \cite{Trotzky295}. They do not play an important role in our experiments.

\begin{figure}[!tb]
\begin{center}
\includegraphics[width=85mm]{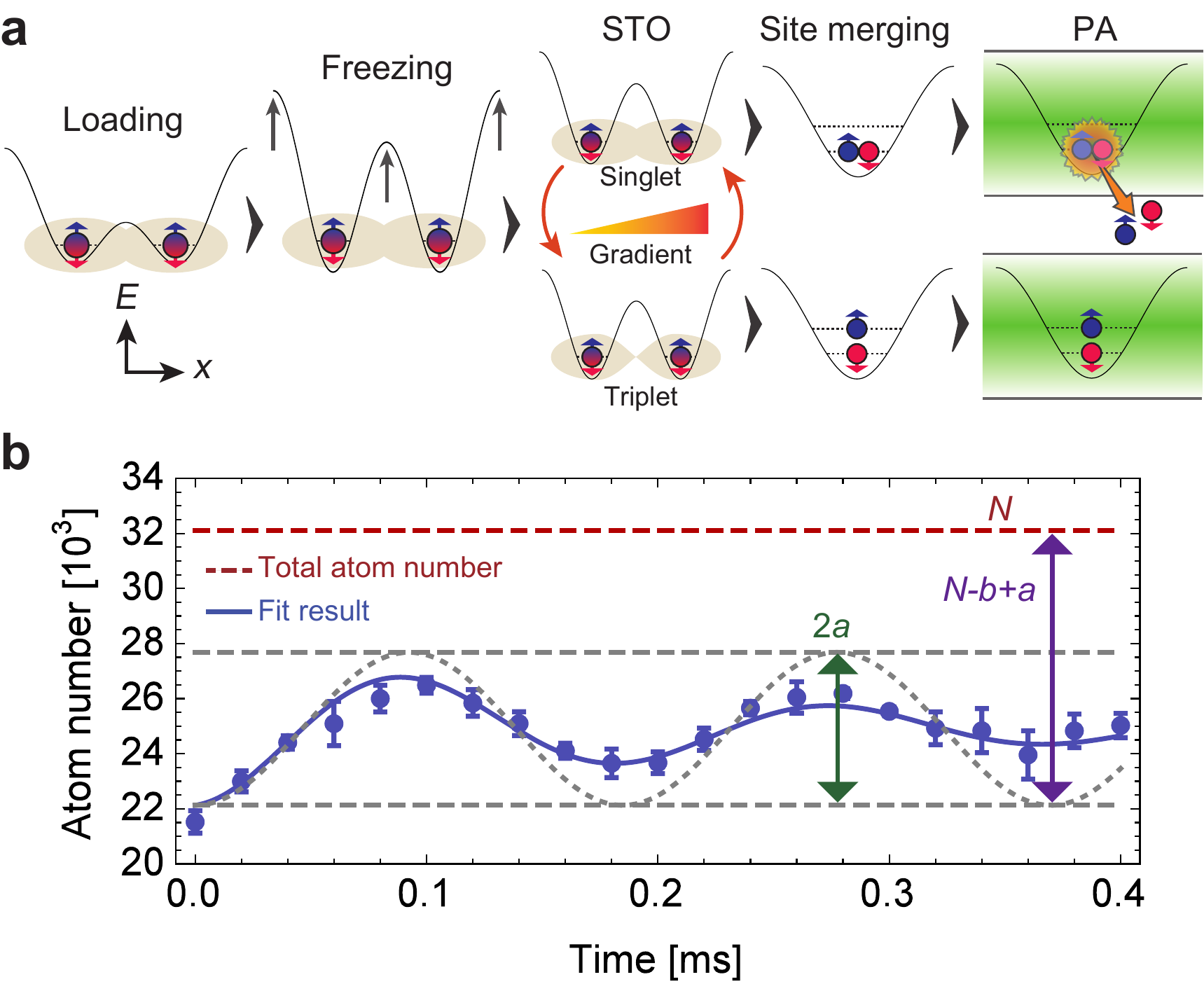}
\caption{(a) Detection sequence for singlets and triplets in a dimer. Shown is the case of two spins (red and blue) per dimer. Depending on the STO time, the two spins form the double occupancy in the lowest band (top), or the state with one spin in the lowest band and the other in the first excited band (bottom) after merging the dimer. These states are distinguished by the PA. (b) Singlet-triplet oscillation in a strongly dimerized lattice for SU(4) spins. The red dashed line represents the total atom number in the lattice without applying the PA. The blue solid line is the fit result with Eq.(\ref{eq:STO}). The gray dotted line is the STO signal assuming no damping. Error bars denote the standard deviation of four independent scans.}
\label{image2}
\end{center}
\end{figure}
%
%
At the early stage of evaporative cooling, we apply the optical pumping \cite{PhysRevLett.105.190401} to create balanced two- or four-component mixtures of $^{173}$Yb (See S.1 in the Supplemental Material (SM) \cite{supp_Lieb} for the details of the optical pumping schemes). The spin distribution after optical pumping is measured by an Optical Stern-Gerlach (OSG) technique \cite{PhysRevLett.105.190401}, where we apply the spin-dependent gradient by an circularly polarized laser beam with a Gaussian profile. After loading the two- or four-component Fermi gas into a strongly dimerized lattice, where all beams are simultaneously ramped in 150 ms with a spline-shaped laser intensity, we detect an antiferromagnetic spin correlation with the sequence as shown in Fig.\ref{image2} (a), similar to Ref.\cite{Greif1307}. In the first part of the detection sequence, we freeze out the atomic motion by applying a two-step ramp of $s_{\rm{L}}=[(20,20.8),48,48] \rightarrow [(25,20.8),80,100] \rightarrow [(25, 100), 80, 100]$. The first ramp and the second ramp take 0.5 ms and 10 ms, respectively. This lattice ramp also removes the contribution of the admixture of double occupancies in the ground-state singlet (S.2 in SM \cite{supp_Lieb}). Then, we apply a spin-dependent gradient by a fictitious magnetic field of light, similar to the OSG beam. This gradient creates an energy difference $\Delta$ for atoms with different spins on neighboring sites and drives coherent oscillation between the singlet$=\left( \ket{\sigma_1, \sigma_2} - \ket{\sigma_2, \sigma_1} \right)/\sqrt{2}$ and triplet $\ket{t_0}=\left( \ket{\sigma_1, \sigma_2} + \ket{\sigma_2, \sigma_1} \right)/\sqrt{2}$ states at a frequency of $\Delta/\hbar$ \cite{PhysRevLett.105.265303}, where $\sigma_i \;(i=1,2)$ denotes a spin component. For a four-component mixture, we use a linearly polarized gradient beam, with which the STOs have the same frequency for the 4 spin pairs of $(m_F=5/2,1/2), (5/2,-1/2), (-5/2,1/2),$ and $(-5/2,-1/2)$, but do not occur for the 2 spin pairs of $(5/2,-5/2)$ and $(1/2,-1/2)$ (S.3 in SM  \cite{supp_Lieb}). After a certain oscillation time, we remove the gradient and merge the dimers into single sites by ramping the lattice potential down to $s_{\rm{L}}=[(25,0),80,100]$ in 1 ms. Due to a fermion anticommutation relation and symmetry of the two-particle wave function, the singlet state on adjacent sites evolves to a doubly occupied site with both atoms in the lowest band, while the triplet state transforms into a state with one atom in the lowest band and the other in the first excited band. The fraction of atoms forming double occupancies in the lowest band is detected by a photoassociation (PA) technique \cite{Taie2012, Sugawa2011, PhysRevLett.93.073002}. The PA process enables us to convert all atoms forming double occupancies in the lowest band into electronically excited molecules that rapidly escape from the trap, whereas the state with one atom in the lowest band and the other in the first excited band is not converted due to its odd-parity of relative spatial wave functions \cite{RevModPhys.78.483}. Therefore, the loss of atoms corresponds to the number of atoms forming the singlet state in the initial dimerized lattice. We note that the symmetric and antisymmetric states$=(\ket{\sigma_1, 0} \pm \ket{0, \sigma_1})/\sqrt{2}$ also exist, especially in the trap edge, but they evolve to the state with one atom per site after merging the dimer, which is not detected by a PA. The PA laser is detuned by -812.26 MHz from the $^1S_0 \leftrightarrow ^3$$P_1(F^{\prime}=7/2)$ transition and has sufficient intensity to finish removing double occupancies within 0.5 ms irradiation.

Figure \ref{image2} (b) shows the typical STO of SU(4) spins in a strongly dimerized lattice. A clear oscillation is visible. The damping of oscillation is caused by the spatial inhomogeneity of the fictitious magnetic gradient and the photon scattering from the gradient beam (S.4 in SM \cite{supp_Lieb}). This oscillation reveals an excess number of singlets compared to triplets, corresponding to an antiferromagnetic correlation on neighboring sites. An STO signal is also observed for an SU(2) system. We fit the data with the empirical function
\begin{equation}
\label{eq:STO}
F(t_{\rm{STO}})= -a \; e^{-t_{\rm{STO}} / \tau} {\rm{cos}} \left( 2 \pi f t_{\rm{STO}}  \right) +b,
\end{equation}
where $a,b,\tau, f$ are fitting parameters. Along with the data of STO, we measure the total atom number in the optical lattice without applying the PA laser, $N$. We quantify this correlation by the normalized STO amplitude $A$ and singlet fractions $p_{\rm{s}}$:
\begin{eqnarray}
A&=&
\begin{cases}
2a/N & \text{for SU(2)} \\
3a/N & \text{for SU(4)}
\end{cases}
\label{eq:STOamp} \\
p_{\rm{s}}&=&1-\frac{b-a}{N}.
\end{eqnarray}
We note that the extracted $N-b-a$ exactly corresponds to the actual atom number in the triplet state $\ket{t_0}$ for SU(2) spins, but that is not the case for SU(4) spins because a coherent oscillation does not occur for the spin pairs of $(m_F=1/2, -1/2)$ and $(5/2, -5/2)$. To take this effect into consideration, we compensate the measured STO amplitude by multiplying $3/2$ for SU(4) case as in Eq.(\ref{eq:STOamp}).
\begin{figure}[!tb]
\begin{center}
\includegraphics[width=85mm]{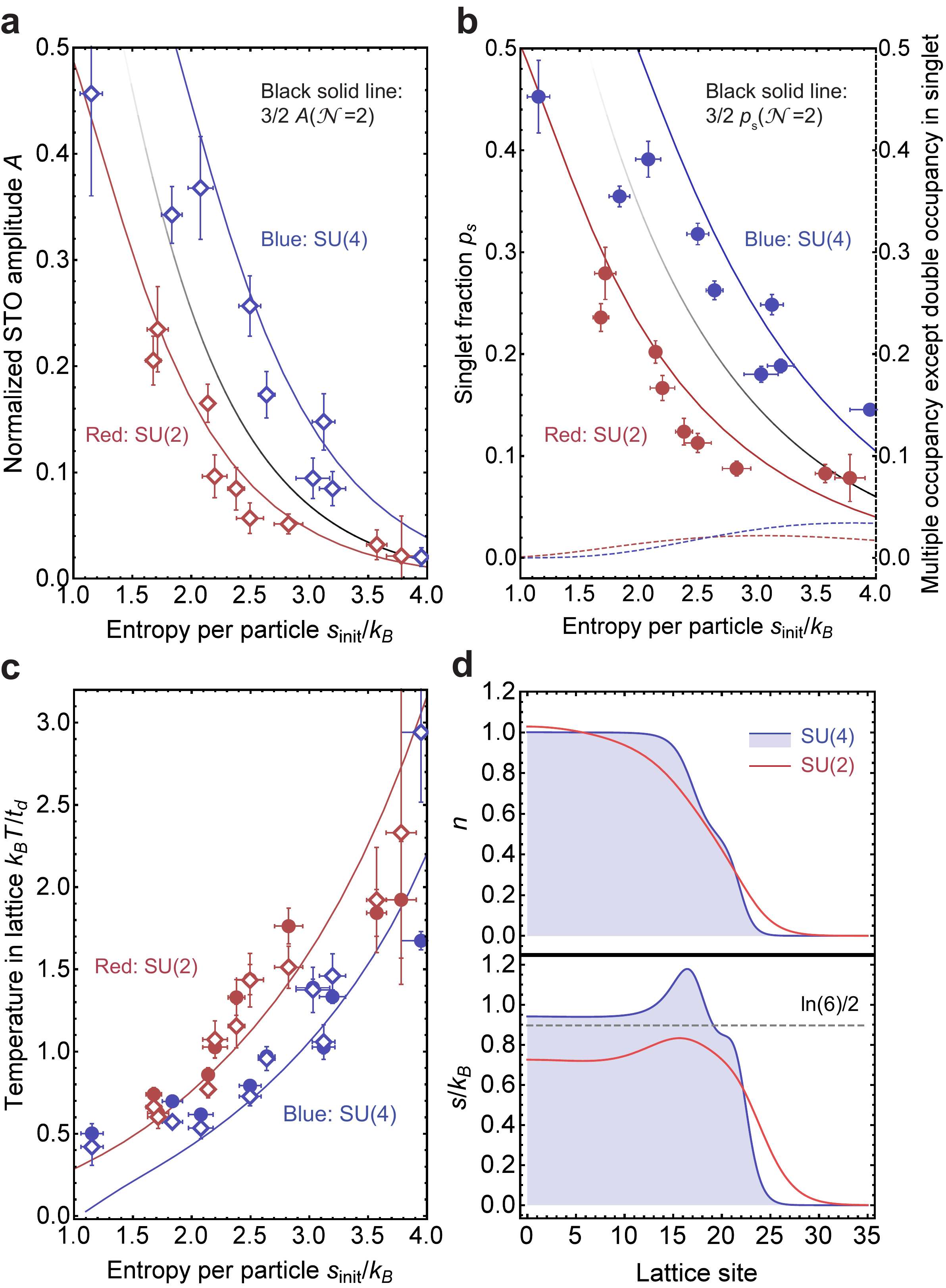}
\caption{(a) Normalized STO amplitude and (b) singlet fraction of SU(2) and SU(4) Fermi gases in the strongly dimerized lattice of $t_{\rm{d}}/t=27$. The dependence on the initial entropy in the harmonic trap is shown. The solid line is a theoretical curve that assumes adiabatic loading into the lattice. The dotted line in (b) is the numerically calculated multiple occupancy except the double occupancy in the ground-state singlet wave function. The vertical error bars include the fitting errors in the STO measurement and the standard deviation of the total atom number $N$. The horizontal error bars show the standard deviation of the 10 independent temperature measurements. (c) Temperature of SU(2) and SU(4) Fermi gases in the lattice. The empty diamond and filled circle are the experimental data estimated from (a) and (b), respectively. Solid line is a theoretical curve. (d) Calculated density (top) and entropy distribution (bottom) at the initial entropy per particle $s_{\rm{init}}/k_{\rm{B}}=1.5$ for SU(2) and SU(4) cases. The maximum singlet entropy per site ln$(6)/2$ for SU(4) is indicated by the gray dashed line.}
\label{image4}
\end{center}
\end{figure}
%
%

To reveal the influence of spin degrees of freedom on the magnetic correlations, we investigate $A$ and $p_{\rm{s}}$ for various initial entropies in the harmonic trap. Figure \ref{image4} (a) and (b) show the results comparing SU(2) and SU(4) systems in a strongly dimerized lattice. The initial temperature in the harmonic trap is obtained by performing the Thomas-Fermi fitting to the 10 independent momentum distributions and the initial entropy $s_{\rm{init}}$ is calculated from the $T/T_{\rm{F}}$ using the formula for a non-interacting Fermi gas, where $T_{\rm{F}}$ is the Fermi temperature. The STO data are taken for the atom number of $N=3.2 \times 10^4$ and the trap frequencies of $(\omega_x, \omega_y, \omega_z)/2\pi = (158.3, 48.6, 141.8)$ Hz, where the filling $n$, i.e., the number of the particle per site, amounts to $n=1$ around the trap center (Fig.\ref{image4} (d)). The solid lines are the result of the atomic limit calculation based on the SU($\mathcal{N}$) FHM in Eq.(\ref{eq:FHM}), assuming the local density approximation (S.5 in SM \cite{supp_Lieb}). The normalized STO amplitude and the absolute singlet fraction decrease for larger entropies, as triplet states become thermally populated. A clear and striking difference between SU(2) and SU(4) systems is visible: the antiferromagnetic correlation is enhanced in the SU(4) system compared to SU(2) for the same initial entropy. 
There are two effects at play. One is the difference of the fraction of singlet configurations among all possible states. The other is the thermodynamic cooling effect related to spin entropy. To discuss these effects, we consider  two atoms with SU($\mathcal{N}$) spin symmetry in an isolated dimer, neglecting double occupancies. At the zero temperature, the singlet probability is $p_{\rm{s}}(\mathcal{N})=1$ regardless of $\mathcal{N}$ because the singlet is the ground state. On the other hand, at the infinite temperature, the singlet probability is $p_{\rm{s}}({\mathcal{N}})= W ({\mathcal{N}}) / {\mathcal{N}}^2$ because the probability is determined by the number of the singlet configurations $W(\mathcal{N})=_{\mathcal{N}} \!\!{\rm{C}}_2$. For SU(2) and SU(4), the ratio of the singlet probabilities at the same temperature becomes
\begin{eqnarray}
p_{\rm{s}}(4)/ p_{\rm{s}}(2)&=&
\begin{cases}
1 & \text{for } T=0  \\
3/2 & \text{for } T=\infty .
\end{cases}
\label{eq:singletfracratio}
\end{eqnarray}
Because this ratio monotonically decreases from 3/2 to 1 as the temperature gets lowered, $p_{\rm{s}}(2) < p_{\rm{s}}(4) < 3/2 p_{\rm{s}}(2)$ holds for a finite temperature. The same inequality is true for $A(4)$ and $A(2)$. The black solid lines in Fig.\ref{image4} (a) and (b) indicate $3/2 p_{\rm{s}}(2)$ and $3/2 A(2)$, which should give the upper limit for $p_{\rm{s}}(4)$ and $A(4)$ at the same temperature. Most of the observed SU(4) data are above the black lines at the same initial entropy. This means that the temperature of SU(4) is lower than that of SU(2), which is ensured from Fig.\ref{image4} (c). This behavior can be understood as follows. Entropy per site of the singlet ground states is given by ln($W(\mathcal{N})$)/2. In contrast to the zero-entropy ground state of SU(2) system, the SU(4) ground state has a residual entropy of ln$(6)/2=0.9$. Therefore, the initial temperature required for spins to form the singlet is increased in the SU(4) system compared to SU(2). This is closely related to the Pomeranchuk effect \cite{RevModPhys.69.683} enhanced by large spin degrees of freedom, which was already demonstrated in the paramagnetic SU(6) fermionic Mott-insulator \cite{Taie2012}. In this work, it is clearly shown that cooling with large SU($\mathcal{N}$) spin can be applied even in the regime of quantum magnetism. We note that in a trapped system, entropy is stored in a low-density metallic state near the edge of the atomic cloud and singlet states at the trap center survive for higher total entropy as in Fig.\ref{image4} (d). The data in Fig.\ref{image4} (a) and (b), especially at low initial entropies, show the discrepancy with the theory. This might be caused by several reasons including some non-adiabaticity in the lattice loading or an imperfect efficiency on the PA (S.6 in SM \cite{supp_Lieb}). From middle to high initial entropies, the measured singlet fraction is slightly overestimated because the multiply occupied states except the ground-state singlet are thermally populated at the initial lattice depth and detected by PA after merging.
\begin{figure}[!tb]
\begin{center}
\includegraphics[width=85mm]{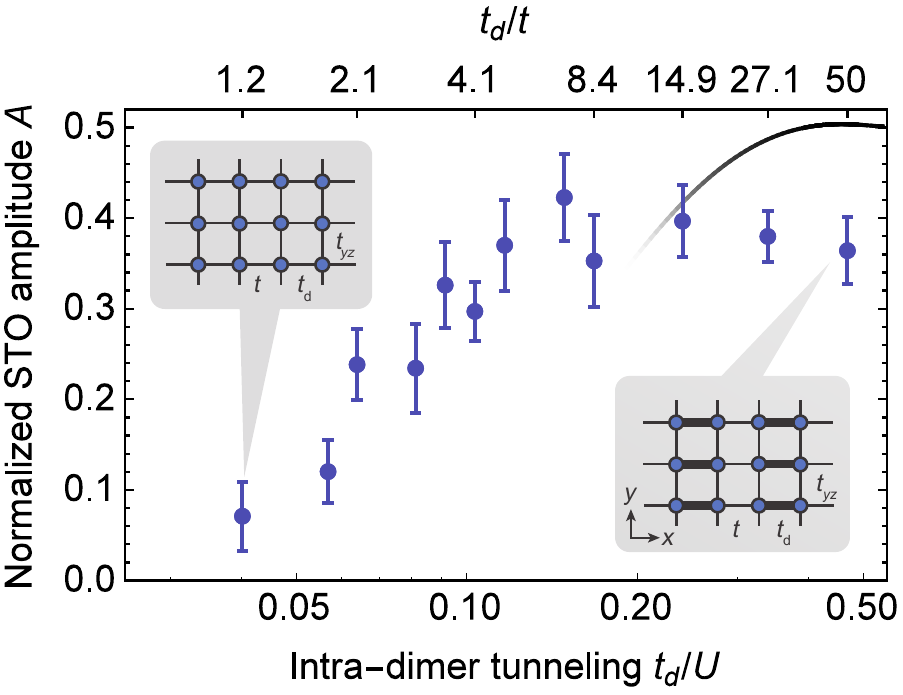}
\caption{Normalized STO amplitude for the SU(4) Fermi gas versus the intra-dimer tunneling. The bottom axis is shown in a logarithmic scale. The black solid curve is the prediction in the atomic limit for an entropy per particle of $s/k_{\rm{B}}=1.9$ under the assumption of the adiabatic loading into the lattice, and is shown down to $t_{\rm{d}}/t=10$. For the entire data, the on-site interaction is fixed to $U/h=3.0$ kHz, while $t$ changes from $t/h=28.0$ Hz to $100$ Hz, and $t_{yz}/t$ from 1.7 to 1.0.}
\label{image5}
\end{center}
\end{figure}

%
%

Finally, we investigate the dependence of the normalized STO amplitude on the intra-dimer tunneling $t_{\rm{d}}$. Figure \ref{image5} shows the result with the SU(4) Fermi gas. The solid line is the theoretical curve shown only for $t_{\rm{d}}/t=10$ and higher. Below this value the atomic limit calculation starts to be invalid. As $t_{\rm{d}}$ decreases, the STO amplitude gets smaller because the excitation energy to the triplet state, which is determined by the exchange energy $-U/2+\sqrt{16t_{\rm{d}}^2+U^2}/2$, is lowered. Our experimental data show such a tendency and indicate the possibility that the nearest-neighbor antiferromagnetic correlation still remains slightly even in the isotropic lattice. In terms of the entropy, the rough criterion for the onset of the nearest-neighbor spin correlation in the lattice is $s/k_{\rm{B}}=$ln$(\mathcal{N})$ \cite{PhysRevLett.109.205306}, which amounts to ln$(\mathcal{N}=4)=1.38$ for SU(4) system. Even though the average entropy in our trapped system is 1.9 in Fig.\ref{image5}, the lower entropy is achieved at the trap center. The atoms around such a region are considered to contribute to the possible nearest-neighbor spin correlation in the isotropic lattice.

%
%
In conclusion, we have studied the important role of the spin degrees of freedom on the antiferromagnetic correlation in a strongly dimerized lattice by comparing the SU(2) and SU(4) systems. We observed the enhanced antiferromagnetic correlation in SU(4) due to the Pomeranchuk effect. Further cooling can be expected for a larger spin system such as SU(6), which $^{173}$Yb possesses. The bottleneck for experiments with higher spin system is the detection technique: if we applied the scheme performed here to SU(6) system of $^{173}$Yb, we would suffer from the multiple STO frequencies. We expect that combining SU($\mathcal{N}>2$) Fermi gas with more complex lattice geometry like a plaquette, which has been already implemented with optical lattices \cite{Dai2017, PhysRevLett.108.205301}, will open up the door to the interesting magnetic order \cite{PhysRevLett.62.1694, PhysRevLett.81.3527}.

\begin{acknowledgments}
This work was supported by the Grant-in-Aid for Scientific Research of MEXT/JSPS  KAKENHI(No. 25220711, No. 26247064, No. 16H00990, No. 16H00801, and No. 16H01053), the Impulsing Paradigm Change through Disruptive Technologies (ImPACT) program, JST CREST(No. JPMJCR1673), and the Matsuo Fundation. H.O. acknowledges support from JSPS Research Fellowships.
\end{acknowledgments}

%

%
%
\pagebreak
\begin{widetext}
\begin{center}
\textbf{\large Supplemental Material for \\ Antiferromagnetic spin correlation of SU($\mathcal{N}$) Fermi gas in an optical super-lattice}
\end{center}
\end{widetext}
\setcounter{equation}{0}
\setcounter{figure}{0}
\setcounter{table}{0}
\setcounter{page}{1}

\makeatletter
\renewcommand{\theequation}{S\arabic{equation}}
\renewcommand{\thefigure}{S\arabic{figure}}
\renewcommand{\refname}{S\arabic{figure}}
\makeatother

%
%
\maketitle
\section{S.1 Optical pumping}

\begin{figure}[!tb]
\begin{center}
\includegraphics[width=85mm]{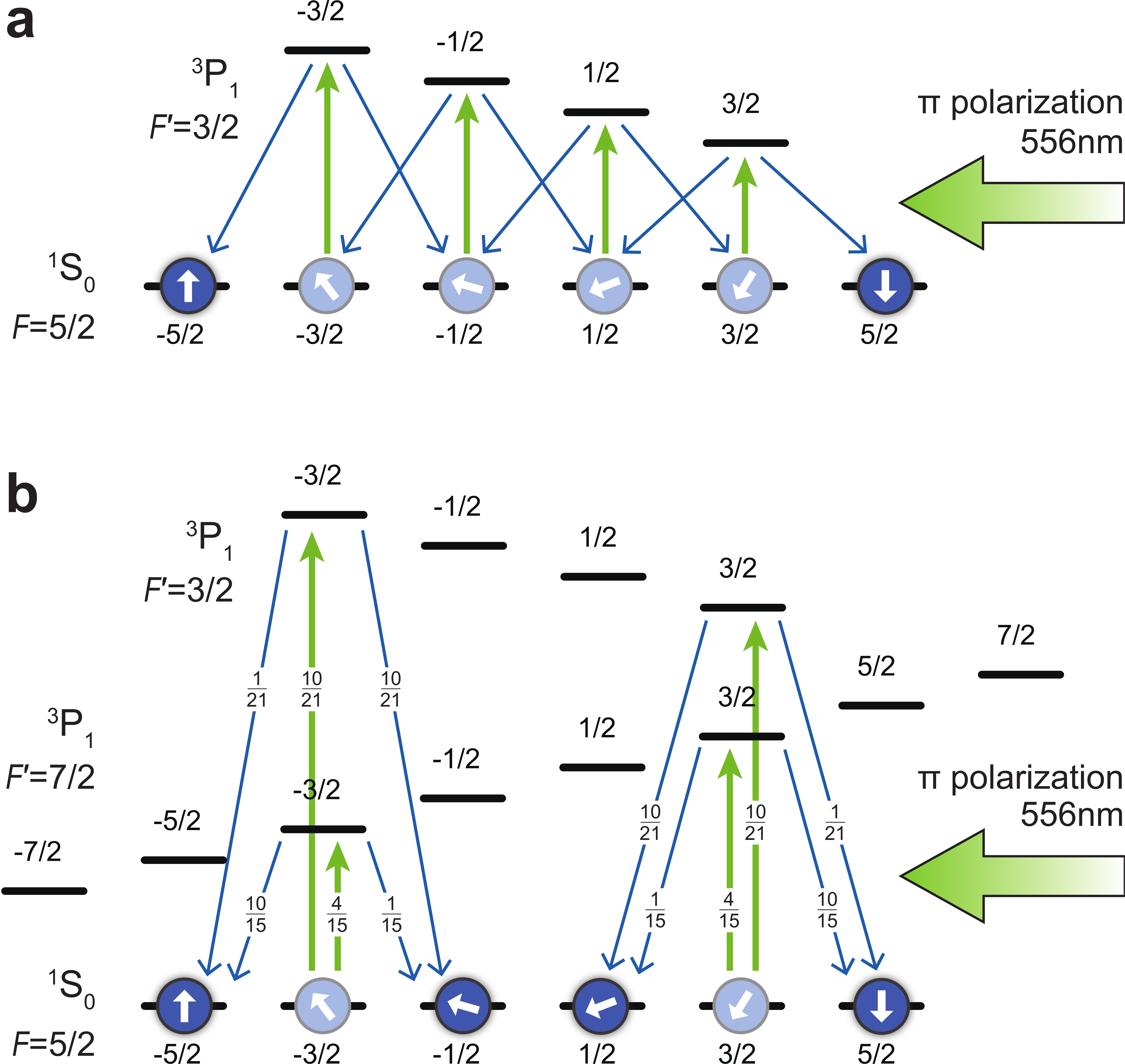}
\caption{Optical pumping schemes to create (a) two- and (b) four-component mixtures of $^{173}$Yb by using a $\pi$-polarized 556 nm light beam. Green and blue arrows mean excitation and decay, respectively. The number labeled on the arrows denotes the absolute square of the normalized Clebsh-Gordan coefficient, which corresponds to the transition strength.}
\label{figureS1}
\end{center}
\end{figure}

Figure \ref{figureS1} illustrates the schematics of optical pumping via the  $^1S_0(F=5/2) \leftrightarrow ^3$$P_1(F^{\prime}=3/2,7/2)$ transitions. For creation of a balanced two-component mixture, we split the sublevels of the $^3$$P_1(F^{\prime}=3/2)$ state by applying an external magnetic field $B\simeq 16$ Gauss, and irradiate the $\pi$-polarized light beam with four resonant frequencies. As a result, an initial six-component ensemble is pumped into two sublevels of $m_F=\pm5/2$ as in Fig.\ref{figureS1} (a). For preparation of a balanced four-component mixture, we use both of the $F^{\prime}=3/2$ and $F^{\prime}=7/2$ states in $^3$$P_1$ in order to compensate the asymmetry of Clebsh-Gordan coeffcients (CGC) \cite{Metcalf1999}. For example, if we are to optically pump atoms in $m_F=3/2$ by using the $\ket{F=5/2, m_F=3/2} \rightarrow \ket{F^{\prime}=7/2, m_{F^{\prime}}=3/2}$ line alone, almost all atoms would be transported to $m_F=1/2$. On the contrary, with the $\ket{F=5/2,m_F=3/2} \rightarrow \ket{F^{\prime}=3/2, m_{F^{\prime}}=3/2}$ line alone, they are pumped into $m_F=5/2$ rather than $m_F=1/2$. Therefore, simultaneous pumping with these transition lines can produce a balanced four-component mixture of $m_F=\pm1/2$ and $m_F=\pm5/2$ as in Fig.\ref{figureS1} (b).

The transition strengths are determined by the absolute square of the CGC, which are given in Fig.\ref{figureS1} (b) for the transitions of our interest. Note that the values are normalized so that the summation over the possible transitions from a particular state in $^3$$P_1$ becomes one.

Figure \ref{image2} shows the observed images of spin distributions after optical pumping. We fit the data with a multi-component Gaussian function to extract the atom number in each spin. The measured spin population is $\left[ p_{m_F=-5/2}, p_{m_F=5/2} \right]=\left[ 0.49(1), 0.50(1) \right]$ in (a), and $\left[ p_{-5/2}, p_{-1/2}, p_{1/2}, p_{5/2} \right]= \left[ 0.26(1), 0.24(1), 0.25(1), 0.25(1) \right]$ in (b). In this way, we successfully create almost balanced two- and four-component mixtures.

%
%
\section{S.2 Detection lattice ramp}
After loading the atoms into the optical lattices, we increase the lattice depth in two steps for measurements. In the first ramp to the point $s_{\rm{L}}=\left[ (25, 20.8), 80, 100 \right]$ in $0.5$ ms, we isolate the individual dimers. In the second ramp to $s_{\rm{L}}=\left[ (25, 100), 80, 100 \right]$ in $10$ ms, we freeze out the atomic motion within a dimer.

Here, we evaluate the adiabaticity of this detection ramp by using a two-component Hamiltonian of two site basis with a time-dependent tunneling $t_{\rm{d}}(\tau)$ and interaction $U(\tau)$ such as
\begin{eqnarray}
H(\tau) = \left(
\begin{array}{cccc}
0 & 0 & -t_{\rm{d}}(\tau) & - t_{\rm{d}}(\tau) \\
0 & 0 & t_{\rm{d}}(\tau) & t_{\rm{d}}(\tau) \\
-t_{\rm{d}}(\tau) & t_{\rm{d}}(\tau) & U(\tau) & 0 \\
-t_{\rm{d}}(\tau) & t_{\rm{d}}(\tau) & 0 & U(\tau) \\
\end{array}
\right),
\label{eq:two-site-ham}
\end{eqnarray}
where we use the localized spin basis of $\left( \ket{\sigma_1, \sigma_2},\; \ket{\sigma_2, \sigma_1},\; \ket{\sigma_1 \sigma_2, 0}, \; \ket{0, \sigma_1 \sigma_2} \right)$.
As an initial state, we use an exact singlet ground state
\begin{eqnarray}
\ket{s(\tau_0)}&=&\frac{-4 t_{\rm{d}}(\tau_0)}{U(\tau_0)-\sqrt{16 t_{\rm{d}}^2(\tau_0) + U(\tau_0)^2}} (\ket{\sigma_1, \sigma_2} - \ket{\sigma_2, \sigma_1})  \nonumber \\
&& + (\ket{\sigma_1 \sigma_2, 0}+\ket{0, \sigma_1 \sigma_2}).
\label{eq:Singlet}
\end{eqnarray}
We numerically solve a time-dependent Schr\"{o}dinger equation
\begin{eqnarray}
i \hbar \frac{d }{d\tau} \ket{\psi (\tau)} = H(\tau) \ket{\psi (\tau)},
\end{eqnarray} 
where we assume a linear change of $U(\tau)$ and $t_{\rm{d}}(\tau)$ in a two step ramp of $\left[U(\tau_0), t_{\rm{d}}(\tau_0)\right]/h=\left[3.0, 1.0 \right]$ kHz $\rightarrow \left[U(\tau_1), t_{\rm{d}}(\tau_1)\right]/h=\left[4.3, 1.3 \right]$ kHz $\rightarrow \left[U(\tau_2), t_{\rm{d}}(\tau_2)\right]/h=\left[7.25, 0.0108 \right]$ kHz. The adiabaticity can be quantified by evaluating the overlap between the time-evolved state $\ket{\psi (\tau_2)}$ and the singlet state $\ket{s(\tau_2)}$ in the final lattice depth. The numerical calculation result is $\left| \braket{s(\tau_2) | \psi(\tau_2)} \right|^2> 0.999$, which ensures the adiabaticity of our detection ramp.

We note that the contribution of the admixture of double occupancies in the singlet state in Eq.(\ref{eq:Singlet}) is removed after the detection lattice ramp due to the large value of $U(\tau_2)/t_{\rm{d}}(\tau_2)$. Therefore, the Fermi-Hubbard singlet state in Eq.(\ref{eq:Singlet}) is converted to a simple singlet state $(\ket{\sigma_1, \sigma_2} - \ket{\sigma_2, \sigma_1})/\sqrt{2}$.
\begin{figure}[tb]
\begin{center}
\includegraphics[width=85mm]{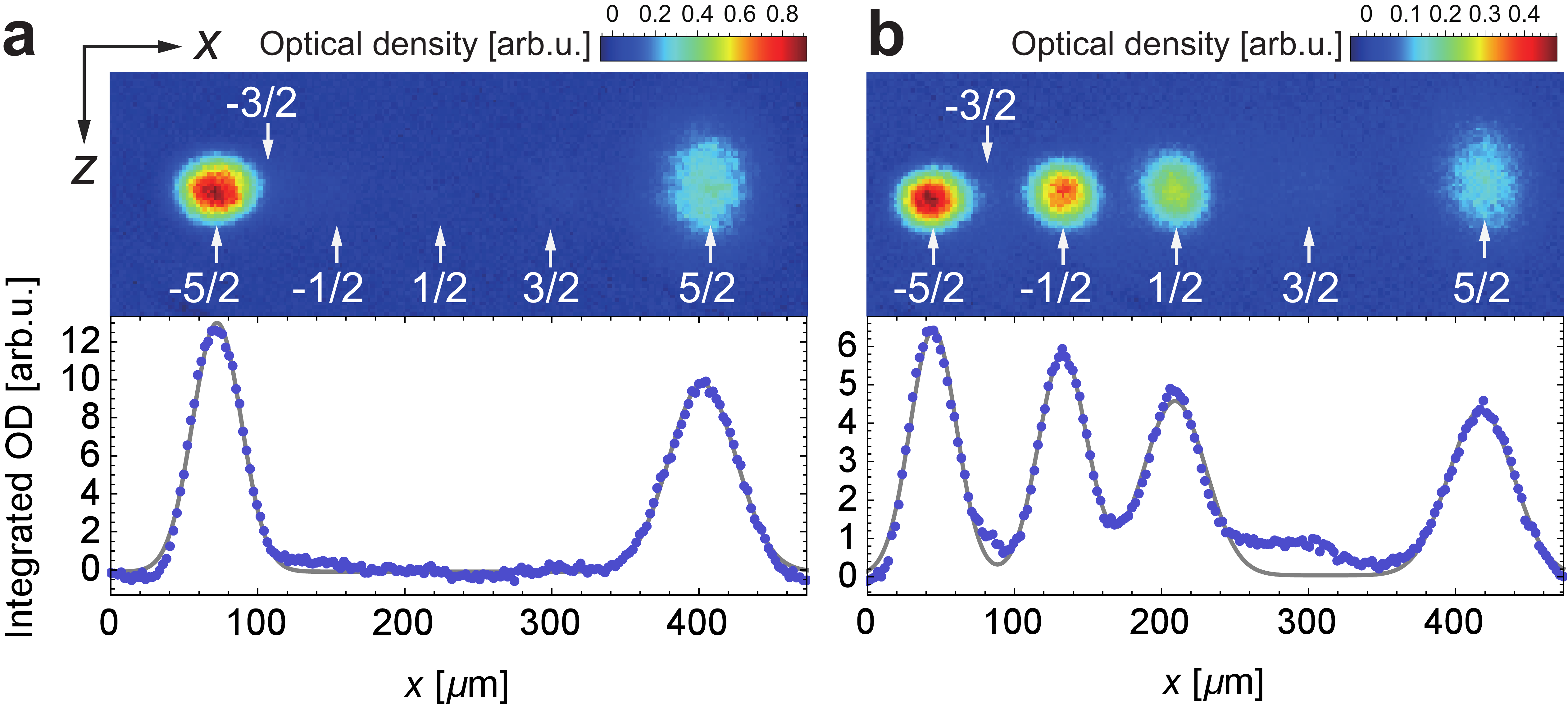}
\caption{Optical Stern-Gerlach separation of spin components after optical pumping to create (a) two- and (b) four-component mixtures. The expansion time is 7 ms. The absorption image is averaged over 10 independent measurements. White arrows point to the atoms in each sublevel of the $^1S_0$ state. The gray line is a fit to the integrated optical density with a multi-component Gaussian function. }
\label{image2}
\end{center}
\end{figure}
%
%
\section{S.3 Light shift of the gradient beam}

In the main text, we use the spin-dependent gradient for the OSG and STO experiments. Here, we calculate the light shift by the gradient beam detuned by a frequency on the order of the hyperfine splitting in $^3$$P_1$ of $^{173}$Yb. The light shift for an atom in $\ket{F,m_F}$ of the lowest-lying $^1S_0$ state is given by  \cite{Cohen-Tannoudji1993}
\begin{flalign}
&\Delta E_{F, m_F} = -\frac{e^2 I_{\rm{laser}}}{ \epsilon_0 \hbar c }  \times& \nonumber \\ 
&  \sum_{F^{\prime},m_{F^{\prime}}} \left[\frac{\omega_{F^{\prime}, m_{F^{\prime}} }}{ \omega_{F^{\prime}, m_{F^{\prime}}}^2 - \omega_{\rm{laser}}^2 } \left|  \braket{F^{\prime}, m_{F^{\prime}} | r_q | F, m_F } \right|^2 \right],& \label{eq:lightshift}
\end{flalign}
where $I_{\rm{laser}}$ is the intensity of laser light, $\omega_{\rm{laser}}$ is the laser frequency, $\omega_{F^{\prime}, m_{F^{\prime}}}$ is the resonant frequency between the $\ket{F, m_F}$ and $\ket{F^{\prime}, m_{F^{\prime}}}$ states, the summation is taken over all the possible $F^{\prime}$ and $m_{F^{\prime}}$, and $\braket{F^{\prime}, m_{F^{\prime}} | r_q | F, m_F }$ is the dipole matrix element defined as
\begin{eqnarray}
\braket{F^{\prime}, m_{F^{\prime}} | r_q | F, m_F } = \mathcal{C}_{s s^{\prime}} (q) \frac{ \braket{\alpha^{\prime} J^{\prime} || r || \alpha J} }{\sqrt{2J^{\prime} +1 }}.
\end{eqnarray}
Here, $ \braket{\alpha^{\prime} J^{\prime} || r || \alpha J} $ is the reduced dipole matrix element, which is specific to the atomic species. In the case of $^1S_0 \leftrightarrow $$^3$$P_1$ of Yb, it amounts to $  \braket{\alpha^{\prime} J^{\prime} || r || \alpha J}  = 0.54 \;a_0$ \cite{PhysRevA.60.2781}, where $a_0$ is the Bohr radius. $\mathcal{C}_{ss^{\prime}}(q)$ is dependent on the quantum numbers in a given transition $\ket{s} \equiv \ket{F(J(LS)I), m_F} \rightarrow \ket{s^{\prime}} \equiv \ket{F^{\prime} (J^{\prime} (L^{\prime} S^{\prime})I^{\prime}), m_{F^{\prime}}}$, and is related to the CGC. This $\mathcal{C}_{ss^{\prime}}(q)$ is given by 
\begin{eqnarray}
\mathcal{C}_{s s^{\prime}}(q) &=& (-1)^{F^{\prime}+J^{\prime}+I^{\prime}-m_{F^{\prime}}+F+1} \delta_{I,I^{\prime} }   \nonumber \\
&&  \sqrt{(2F^{\prime}+1) (2F+1) (2J^{\prime}+1)}  \nonumber \\
&&  \left(
\begin{array}{ccc}
F^{\prime } & 1  & 1 \\
-m_{F^{\prime}} & q & m_F \\
\end{array}
\right) \left\{
\begin{array}{ccc}
F^{\prime} & 1 & F \\
J & I^{\prime} & J^{\prime} \\
\end{array}
\right\},
\end{eqnarray}
where the round bracket and curly bracket arrays represent respectively the Wigner's 3j- and 6j-symbol, and $q$ denotes the polarization of the light: $q=0$ for a linear polarization aligned to the quantization axis, and $q=\pm1$ for right- and left-circular polarizations propagating along the quantization axis. 

\begin{figure}[!tb]
\begin{center}
\includegraphics[width=85mm]{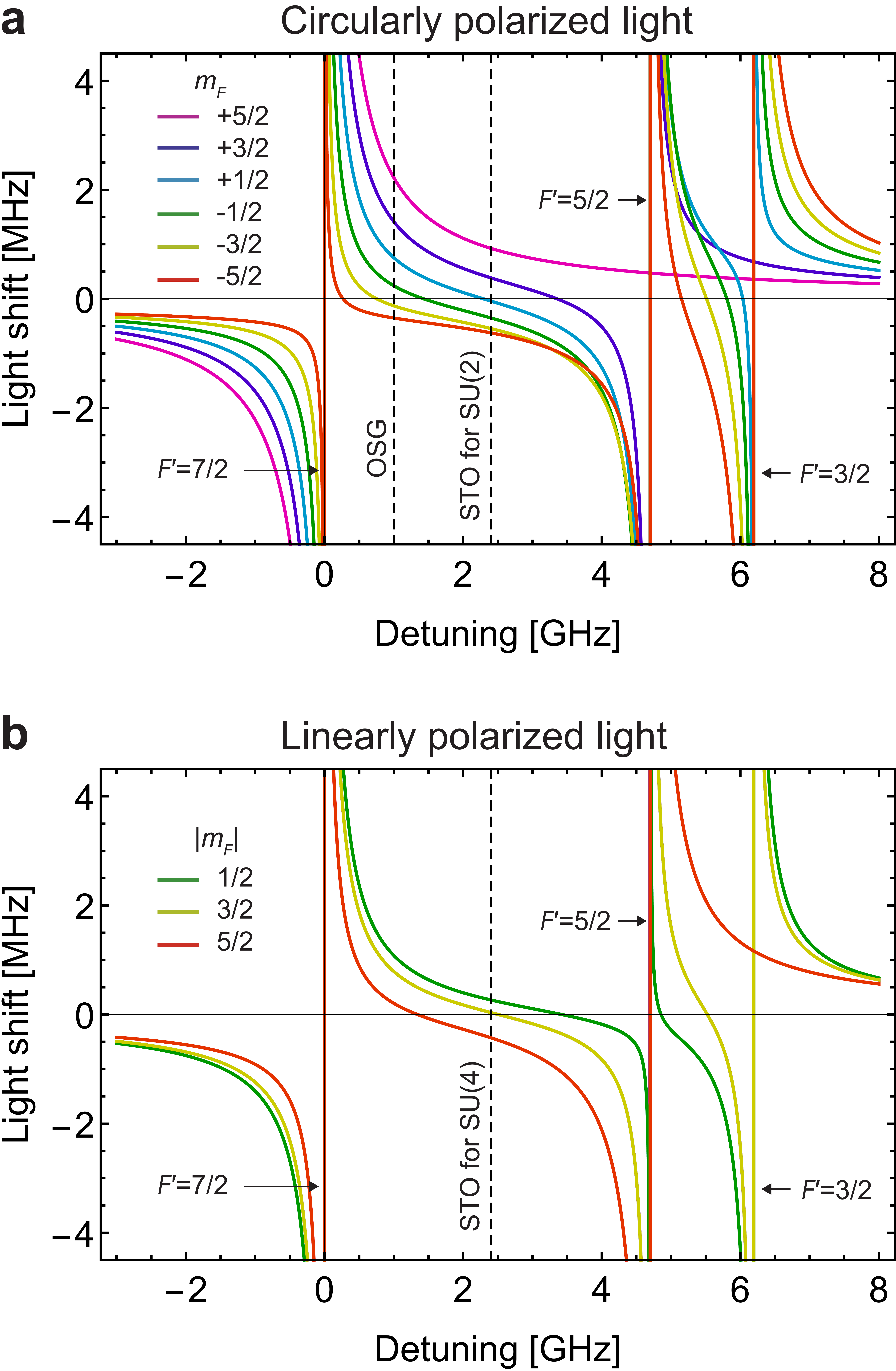}
\caption{Calculated light shifts of (a) a circularly polarized light beam and (b) linearly polarized light beam as a function of frequency detuned from the $^1S_0 \leftrightarrow$$^3$$P_1(F^{\prime}=7/2)$ transition. The vertical dotted line indicates the detuning used for OSG or STO experiments.  We set the laser intensity to $I_{\rm{laser }}=74.1 \;{\rm{W}}/{\rm{cm}}^2$. }
\label{figureS2}
\end{center}
\end{figure}

We take the hyperfine state of $F^{\prime} = 3/2, 5/2, 7/2$ in $^3$$P_1$ into consideration to calculate the Eq.(\ref{eq:lightshift}). Figure \ref{figureS2} (a) and (b) show the calculated light shifts for circular and linear polarizations, respectively. In the OSG experiment, we use the circularly polarized light with 1 GHz blue detuning from the $^1S_0 \leftrightarrow ^3$$P_1(F^{\prime}=7/2)$ transition, with which all six-spin components of $^{173}$Yb can be separated. To induce an STO for a two-spin mixture of $m_F=\pm5/2$, we use a circularly polarized light beam with 2.4 GHz blue detuning, where the associated photon-scattering rate normalized by the light shifts becomes minimum. A linearly polarized light beam with the same detuning is applied for a four-spin mixture of $m_F=\pm1/2$ and $m_F=\pm 5/2$, where the light shifts for an atom with the same absolute value of sublevel are the same.

%
%
\section{S.4 Singlet-triplet oscillation}

We irradiate the spin dependent gradient beam with an isotropic Gaussian profile to the atomic cloud after the detection lattice ramp. The beam waist is $w_0=50$ $\mu$m. We aligned the gradient beam so that the atomic cloud is placed around the steepest gradient point, which is distanced by $w_0/2$ from the beam center. In practice, the STO frequency can be estimated from the separation of the spin distribution by the OSG experiment as in Fig.2 in the main text. The estimated STO frequency is 5.0 kHz, which is consistent with the measured value of $5.1(1)$ kHz in Fig.3 (b).

The simple model to understand the time evolution during STO is obtained by adding the spin-dependent energy offset elements to the Hamiltonian Eq.(\ref{eq:two-site-ham}):
\begin{eqnarray}
H_{\Delta} = \left(
\begin{array}{cccc}
\Delta & 0 & -t_{\rm{d}} & - t_{\rm{d}} \\
0 & -\Delta & t_{\rm{d}} & t_{\rm{d}} \\
-t_{\rm{d}} & t_{\rm{d}} & U+\alpha \Delta & 0 \\
-t_{\rm{d}} & t_{\rm{d}} & 0 & U-\alpha \Delta \\
\end{array}
\right),
\label{eq:ham-grad}
\end{eqnarray}
where $\Delta=(\Delta E_{\sigma_1} - \Delta E_{\sigma_2})/2$ and $\alpha=(\Delta E_{\sigma_1} + \Delta E_{\sigma_2})/(\Delta E_{\sigma_1}-\Delta E_{\sigma_2})$. We define the singlet overlap as
\begin{eqnarray}
O_s \left( t_{\rm{STO}} \right) = \left| \bra{s(\tau_2)} e^{-\frac{i}{\hbar} H_{\Delta} t_{\rm{STO}} } \ket{s(\tau_2)} \right|^2.
\end{eqnarray}
The reliability of the STO can be verified by taking the long term time average of $O_s$
\begin{eqnarray}
\mathcal{A} = \frac{2 \int dt_{\rm{STO}} O_s(t_{\rm{STO}})}{\int dt_{\rm{STO}}}.
\end{eqnarray}
For the STO experiment in the SU(2) system, $\Delta/h=5.0$ kHz and $\alpha=-0.20$. For the 4 spin pairs of $(m_F=\pm5/2, m_F=\pm1/2)$ in the SU(4) system, $\Delta/h=5.0$ kHz and $\alpha=-0.23$. For both cases, the numerical results are $\mathcal{A} > 0.999$. Therefore a clear STO can be expected in a uniform system without any heating effects.

\begin{figure}[!tb]
\begin{center}
\includegraphics[width=85mm]{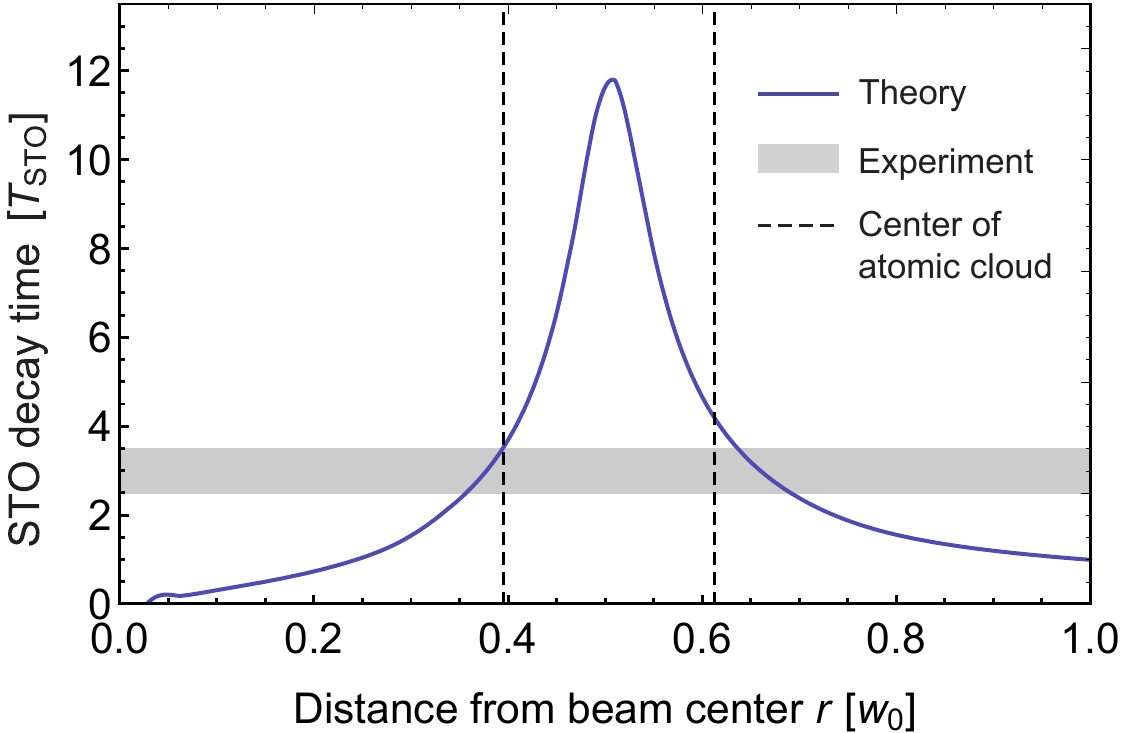}
\caption{Calculated STO decay time caused by the inhomogeneity of the gradient beam. The dependence on the distance from the beam center to the center of atomic cloud is shown. The vertical and horizontal axes are normalized by calculated STO period $T_{\rm{STO}}$ and actual beam waist $w_0$, respectively. We assume the SU(2)-spin system and $w_0=50$ $\mu$m. The shaded region shows the range of the measured values in the STO experiment in SU(2) system. The black dotted line is the center position of the atomic cloud estimated from the OSG experiment.}
\label{figureS3}
\end{center}
\end{figure}

We introduce the spatial dependence to the energy offset $\Delta$ and singlet state $\ket{s}$: $\Delta, \ket{s} \rightarrow \Delta(r), \ket{s(r)}$. The spatial distribution of the singlet state is obtained from the atomic limit calculation of the SU$(\mathcal{N})$ Hubbard model described in S.5. We calculate a global singlet overlap by taking the spatial average of a local singlet overlap $O_s(r,t_{\rm{STO}})$
\begin{eqnarray}
\bar{O}_s\left( t_{\rm{STO}} \right) = \frac{\int dr O_s(r, t_{\rm{STO}})}{\int dr O_s(r, 0)}.
\end{eqnarray}
To evaluate the decay time of STO due to the spatial inhomogeneity of the gradient beam, we make a fitting to $\bar{O}_s (t_{\rm{STO}})$ with a damped cosine function. Figure \ref{figureS3} shows the calculated STO decay time for SU(2) system as the parameter of relative distance from the beam center to the center of atomic cloud. The decay time strongly depends on the relative distance. Even though it is difficult to directly measure the relative distance in our experimental setup, we can indirectly estimate it from the potential gradient measured by the OSG experiment. The black dotted lines are the estimated relative distances, which are displaced from the steepest gradient point $w_0/2$ by about $w_0/10$ due to misalignment. At those points, the experimental STO decay time roughly agrees with the theory. The residual discrepancy is probably due to the photon scattering from the gradient beam, which disturbs the spin balance. The photon scattering rate is given by
\begin{flalign}
&\gamma_{\rm{sc}}=\frac{3\pi c^2 w_{\rm{laser}}^3 I_{\rm{laser}}}{2\hbar} \times& \nonumber \\
& \qquad \sum_{\beta} \left[ \frac{\Gamma_{\beta}}{\omega_{\beta}^3} \left( \frac{1}{\omega_{\rm{laser}} - \omega_{\beta}} + \frac{1}{\omega_{\rm{laser}}+\omega_{\beta}} \right) \right]^2,&
\end{flalign}
where the summation is taken over the possible transitions, $\Gamma_{\beta}$ and $\omega_{\beta}$ are the line width and resonant frequency for each transition, respectively. The other notations are the same as in Eq.(\ref{eq:lightshift}). In practice, we take into account the $^{1}S_0$$ \leftrightarrow $$^{1}P_1$ and $^{1}S_0$$ \leftrightarrow $$^{3}P_1$ transitions. With the parameters used in the STO experiments, the photon scattering rate by the gradient beam is estimated as $\gamma_{\rm{sc}}  \simeq 366$ Hz.

%
%
\section{S.5 Theoretical model}

The theoretical calculations in the main text are based on the atomic limit of the SU($\mathcal{N}$) Hubbard model in a dimerized lattice \cite{Greif1307}. The atomic limit calculation is expected to converge in the parameter regime $t,t_{\rm{yz}} \ll k_{\rm{B}} T \ll U, t_{\rm{d}}$. The evaluation of the grand canonical potential per dimer $\Omega^{\rm{d}}$ in the atomic limit requires the calculation of the isolated dimer Hamiltonian $\hat{H}_{\rm{0}}$ defined in the main text:
\begin{eqnarray}
-\beta \Omega^{\rm{d}} &=& {\rm{log}} z_0^{\rm{d}}, \\
z_0 &=& {\rm{Tr}} \left\{ e^{-\beta \hat{H}_0 }  \right\},
\end{eqnarray}
where $\beta = 1/(k_{\rm{B}}T)$. The trace is evaluated with the localized spin basis containing $2^{2 \mathcal{N}}$ states. Thermodynamic quantities such as the density $n$ and entropy per site $s$ can be obtained from derivatives of the grand canonical potential
\begin{eqnarray}
n &=& -\frac{\partial \Omega^{\rm{d}}}{ \partial \mu}, \\
s &=& -\frac{\partial \Omega^{\rm{d}}}{ \partial T}.
\end{eqnarray}
The evaluation of observables $\hat{\mathcal{O}}$ on a dimer such as the singlet and triplet probabilities is made by calculating the expression
\begin{eqnarray}
\left< \hat{\mathcal{O}} \right> = \frac{{\rm{Tr}} \left\{  \hat{\mathcal{O}} e^{-\beta \hat{H}_0} \right\} }{z_0}.
\end{eqnarray}
The above formalism can be applied to a uniform system. For a trapped system, the effect of the harmonic trap can be included with a local density approximation (LDA), where we assume that the density of the atoms in an optical lattice smoothly changes. The harmonic confinement leads to a quadratically varying chemical potential
\begin{eqnarray}
\mu \rightarrow \mu(r) = \mu_0 - \frac{1}{2} m \bar{\omega}^2 \left( \frac{\lambda }{2} \right)^2 r^2,
\end{eqnarray}
where $\bar{\omega}=(\omega_x \omega_y \omega_z)^{1/3}$ is the geometric mean of the trapping frequencies, $\mu_0$ is the chemical potential at the trap center, and $r$ is the normalized distance from the trap center to a site. The averaged observable $\left< \hat{\mathcal{O}}^{\rm{LDA}} \right>$ can be obtained by integrating the contributions per site $\left< \hat{\mathcal{O}}(\mu(r)) \right>$:
\begin{eqnarray}
\left< \hat{\mathcal{O}}^{\rm{LDA}} \right> = \int^{\infty}_0 4 \pi r^2 \left< \hat{\mathcal{O}}(\mu(r)) \right> {\rm{d}}r.
\end{eqnarray}
In the cold atom experiment, the atom number $N$ and entropy per particle $S/N$ in the entire trapped system are accessible quantities. From these numbers, we can determine the system temperature $T$ and chemical potential $\mu_0$ by numerically solving the following simultaneous equations:
\begin{eqnarray}
N &=& \int^{\infty}_0 4 \pi r^2 n(\mu(r) ,T) {\rm{d}}r, \\
S &=& \int^{\infty}_0 4 \pi r^2 s(\mu(r), T) {\rm{d}}r.
\end{eqnarray}

%
%

\section{S.6 Singlet-Triplet imbalance}

\begin{figure}[!tb]
\begin{center}
\includegraphics[width=85mm]{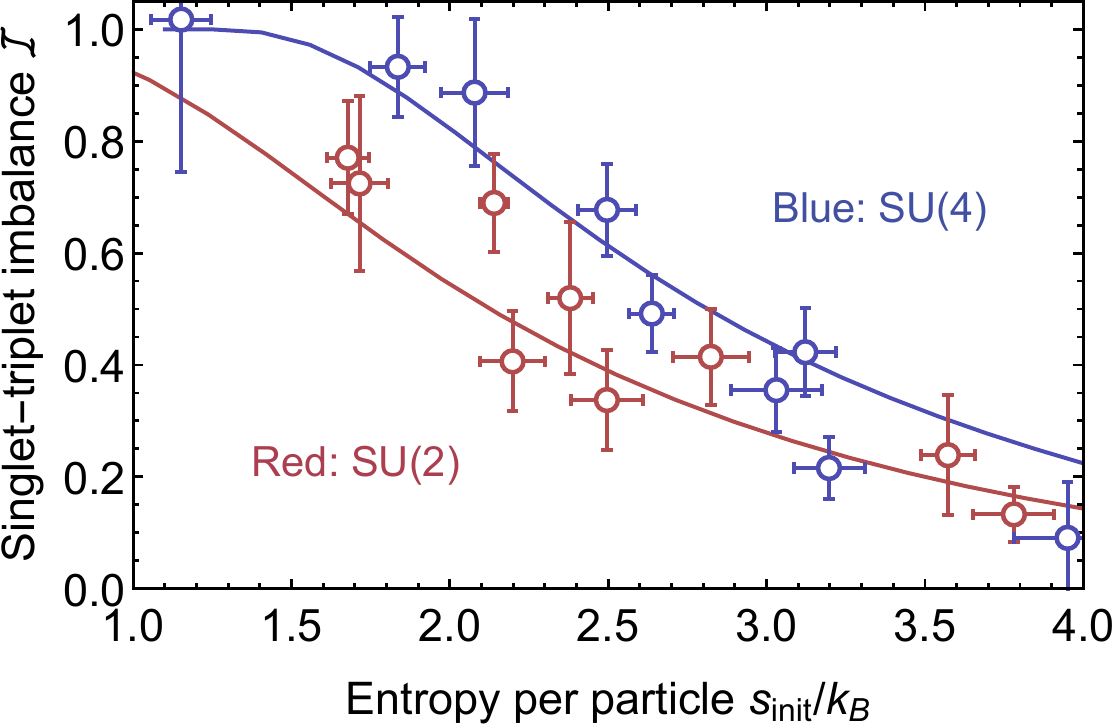}
\caption{The initial entropy in the harmonic trap versus the singlet-triplet imbalance of SU(2) and SU(4) Fermi gases in the strongly dimerized lattice of $t_{\rm{d}}/t=27$. The solid line is a theoretical curve.}
\label{figureS4}
\end{center}
\end{figure}

We can quantify the measured spin correlation by the singlet-triplet imbalance (STI) used in Ref.\cite{Greif1307}:
\begin{eqnarray}
\mathcal{I}&=&\frac{p_{\rm{s}}-p_{\rm{t_0}}}{p_{\rm{s}}+p_{\rm{t_0}}},
\label{eq:STI}
\end{eqnarray}
where $p_{\rm{s}}$ and $p_{\rm{t_0}}$ are the singlet fraction and triplet fraction, respectively. $p_{\rm{s}}$ is defined by Eq.(7) in the main text. The triplet fraction can be extracted by $p_{\rm{t_0}}=1-(b+a)/N$ for SU(2) spins, but that is not the case for SU(4) spins because STOs do not occur for the 2 spin pairs of $(m_F=5/2, m_F=-5/2)$ and $(1/2,-1/2)$. For SU(4) case, the value $1-(b+a)/N \equiv \tilde{p}_{\rm{t_0}}$ is related to the actual triplet fraction $p_{\rm{t_0}}$ and singlet fraction $p_{\rm{s}}$ as
\begin{eqnarray}
\tilde{p}_{\rm{t_0}} &=& \frac{2}{3} p_{\rm{t_0}} + \frac{1}{3} p_{\rm{s}}.
\end{eqnarray}
Substituting $p_{\rm{t_0}}$ for $\tilde{p}_{\rm{t_0}}$ and $p_{\rm{s}}$ in Eq.(\ref{eq:STI}), we obtain the STI for SU(4):
\begin{eqnarray}
\mathcal{I}&=&\frac{p_{\rm{s}} - \tilde{p}_{\rm{t_0}}}{p_{\rm{s}}/3 + \tilde{p}_{\rm{t_0}}}.
\label{eq:STI4}
\end{eqnarray}

Figure \ref{figureS4} shows the STI of SU(2)- and SU(4)-spin systems in a strongly dimerized lattice. The experimental data used for analysis are the same as for Fig.4 (a) and (b) in the main text. The relative error $\delta \mathcal{I}/\mathcal{I}$ is larger than $\delta A/A$ in Fig.4 (a). This can be explained as follows. In our experiments, shot-to-shot fluctuation of the total atom number $N$ is the dominant cause of the error: $\delta N \gg \delta a , \delta b$. From Eq.(\ref{eq:STI}), the relative error of STI can be expressed as
\begin{eqnarray}
\frac{\delta \mathcal{I}}{ \mathcal{I} } &=& \frac{\delta N }{N-b}.
\label{eq:STIerror}
\end{eqnarray}
Assuming that the temperature is sufficiently low, we can use the approximation such as $N-b \approx a$.
Thus, the relative error in Eq.(\ref{eq:STIerror}) becomes
\begin{eqnarray}
\frac{\delta \mathcal{I}}{ \mathcal{I} } &\approx& \frac{\delta N}{a}.
\end{eqnarray} 
In our experiment, the relative error of the total atom number is typically $\delta N /N \approx 5$ $ \% $, and the STO amplitude is $a/N \approx 15 $ $\% $ at most. Therefore, $\delta \mathcal{I}/ \mathcal{I} \approx 33$ $ \% $. In the case of the normalized STO amplitude adapted in the main text, the similar calculation gives
\begin{eqnarray}
\frac{\delta A}{A} &\approx& \frac{\delta N}{N}.
\end{eqnarray}
The above means that the normalized STO amplitude is more robust against the shot-to-shot fluctuation of $N$ than the STI.

Although the uncertainty is large, it seems that the spin correlation of SU(4) in Fig.\ref{figureS4} shows good agreement with the theory even at the low initial entropies, while that is not the case in Fig.4 (a) and (b). Unlike $A$ and $p_{\rm{s}}$ defined in the main text, the STI is sensitive only to dimers containing two atoms and does not include the information of low-density states with one atom per dimer.  Therefore, we suspect that the behavior of $A$ and $p_{\rm{s}}$ could be attributed to the unexpectedly large fraction of the low-density states, which can be caused by several reasons:
\begin{enumerate}
\item Non-adiabaticity of lattice loading \\
	For a strongly dimerized lattice with large intra-dimer exchange interaction, it is difficult to ensure the overall density profile to be thermally equilibrated rather than the local STI. Melting of the density plateau shown in Fig.4 (d) results in the increase of the low-density states.
\item Three-body loss \\
	For the SU(4) case, three-body loss during the lattice loading process takes place. It reduces the density, especially in the region near the trap center.
\end{enumerate}
In addition, although we checked that the PA detection works properly for all possible spin combinations, imperfection in detecting doubly occupied states after merging results in the overestimation of the low-density states.

\end{document}